\documentclass[twocolumn,prl,showpacs,superscriptaddress]{revtex4}
\usepackage{epsfig}
\newcommand{\erfc}{\mathrm{erfc}}
\newcommand{\timescale}{\tau_0}
\newcommand{\tangential}{s}

\begin{document}

\title{Crack growth by surface diffusion in viscoelastic media}

\author{R. Spatschek}
\affiliation{Institut f\"ur Festk\"orperforschung, Forschungszentrum J\"ulich, D-52425 J\"ulich, Germany}
\affiliation{Physics Department and Center for Interdisciplinary Research on Complex Systems, Northeastern University, Boston, MA 02115, USA}
\author{E. A. Brener}
\affiliation{Institut f\"ur Festk\"orperforschung, Forschungszentrum J\"ulich, D-52425 J\"ulich, Germany}
\author{D. Pilipenko}
\affiliation{Institut f\"ur Festk\"orperforschung, Forschungszentrum J\"ulich, D-52425 J\"ulich, Germany}


\date{\today}

\begin{abstract}
We discuss steady state crack growth 
%in viscoelastic media 
in the spirit of a free boundary problem. 
%a pattern formation process.
It turns out that mode I and mode III situations are very different from each other:
In particular, mode III exhibits a pronounced transition towards unstable crack growth at higher driving forces, and the behavior close to the Griffith point is determined entirely through crack surface dissipation, whereas in mode I the fracture energy is renormalized due to a remaining finite viscous dissipation.
Intermediate mixed-mode scenarios allow steady state crack growth with higher velocities, leading to the conjecture that mode I cracks can be unstable with respect to a rotation of the crack front line.
\end{abstract}

\maketitle

The growth of cracks is major puzzle in solid state physics and materials science, still lacking a convincing physical description.
It links macroscopic material properties to microscopic effects in the tiny tip region and raises the important question which features of crack growth are generic and can be attributed to larger classes of materials.
%The multiscale modeling of crack growth is therefore currently attacked from very different approaches ranging from quantum-mechanical calculations of the tip region, atomistic molecular dynamics simulations to fully macroscopic descriptions.
The latter has recently attracted interest through the phase field method, which reformulates crack propagation as a moving boundary problem, where the tip scale can be either determined through intrinsic scales of the modeling technique or by macroscopic selection principles.
Here, the Asaro-Tiller-Grinfeld instability (ATG) \cite{Asaro,Grinfeld}, which is usually understood as the morphological instability of a uniaxially stressed surface or interface, turns our to be strongly related to this problem:
A perturbation of an initially straight surface increases the surface area, but diminishes the stored elastic energy, and therefore decreases the total energy of the system, provided that the wavelength of perturbation exceeds a critical (macroscopic) lengthscale, which depends on the applied stress.
We note that the same ingredients, a counterplay between a release of elastic energy and an increase of surface or fracture energy are the basic mechanisms to understand the propagation of cracks beyond the Griffith point.
%The evolution of the surface is usually driven by surface diffusion, or, in the case of a solid phase in contact with its melt, by melting and solidification.
Nevertheless, it is known that this instability leads to a breakdown of the physical description, as the unstable solid forms deep grooves, which, after a finite time, advance with infinitely high velocity and vanishing tip radius.
The reason for this breakdown is the absence of an additional microscopic lengthscale for selection of a crack-like tip radius.
Hence, understanding the selection of a crack tip radius in fracture has important implications also for the stability of stressed surfaces.
%We also demonstrated that the inclusion of inertial effects, which limits the propagation velocity to values appreciably below the Rayleigh speed, indeed induces such a selection mechanism;
%for high applied tensions a branching instability can occur.

%The assumption of small-scale yielding is often used in modeling of cracks, which means that all dissipation takes place in a very narrow region around the tip.

One of the central questions for any crack model is the role of dissipation, which is directly connected to the quest for selection mechanisms for a tip scale.
The elastic loading, which is applied far away from the crack tip, is usually only partially used to create the (macroscopically visible) crack surfaces;
especially for higher propagation speeds a microbranching instability can significantly increase the fracture energy \cite{Fineberg}.
This already indicates that the three-dimensional geometry is important for a full understanding of crack propagation, and it is one the goals of this paper to shed light on this by investigating different modes of loading.
In the tip region, the local temperature can rise significantly and even exceed the equilibrium melting temperature, which shows that dissipation can be very strong there.
Although we have demonstrated that even pure dynamical linear elasticity can regularize the singular crack tip \cite{Spatschek06}, it is natural to assume that deviations from a pure elastic behavior can play a crucial role, which can also contribute to dissipation;
plasticity is an important example \cite{Procaccia}, but still a full description has not yet been archived there.
It is obvious that a very detailed modeling of the tip region is required to investigate these different effects, which include a self-consistent selection of the crack shape itself.

To address these important questions, we propose a description of crack propagation in the spirit of interfacial pattern formation processes by inclusion of viscoelastic effects.
This picture goes beyond the usual small scale yielding that is frequently used in the modeling of brittle fracture and includes two dissipative mechanisms:
First, there is dissipation directly at the crack surface;
the incoming flow of elastic energy is partially converted to surface energy in order to advance the crack, and the remaining part is converted to heat.
Second, an extended zone of viscous dissipation is formed around the crack.
We note that this problem is quite complicated as the shape of the crack, its velocity and the distribution between viscous and interfacial dissipation have to be determined self-consistently.

Viscous dissipation in mode I fracture has been discussed in the literature, and although our results qualitatively agree, our model makes a further step as it introduces this effect as way to intrinsically regularize the tip-singularity by selection of the crack tip radius.
In contrast, other models assume a Barenblatt crack tip model or similar ad-hoc regularization criteria.
For details see, for example, \cite{Persson} and references therein.

%The viscous dissipation for Mode I crack has been intensively discussed in the literature . While our results for mode I are in qualitative agreement with existing results, they are quite different in details.
%The point is that account of viscous dissipation requires regularization of the tip singularity which is in our case is provided by self-consistent shape selectiondue to surface diffusion. 
%In many previous papers, the Barenblatt  crack-tip model has been used  and also some other ad-hoc regularization criteria.
%For details see, for example,  B. N. J. Persson and E. A. Brener, PRE 71, 036123. and references therein.

%\section{Model equations}
%{\em Model.}
%
For simplicity we assume that the system obeys a translational invariance in one direction, thus it is effectively two-dimensional.
We assume an isotropic linear viscoelastic medium, $u_i$ and $\epsilon_{ik}$ are displacement and strain respectively.
The total stress, $\sigma_{ik}=\sigma_{ik}^{(el)}+\sigma_{ik}^{(vis)}$, is decomposed into the elastic stress, which is given by Hooke's law (with elastic modulus $E$, Poisson ratio $\nu$),
\begin{equation} \label{model::eq1}
\sigma_{ik}^{(el)} = \frac{E}{1+\nu} \left[ \epsilon_{ik} + \frac{\nu}{1-2\nu} \delta_{ik}\epsilon_{ll} \right],
\end{equation}
and the viscoelastic stress \cite{Landau}
\begin{equation} \label{model::eq2}
\sigma_{ik}^{(vis)} = 2\eta \left[ \dot{\epsilon}_{ik} - \frac{1}{3}\delta_{ik}\dot{\epsilon}_{ll} \right] + \zeta \dot{\epsilon}_{ll} \delta_{ik},
\end{equation}
which is related to the displacement rate through the viscosities $\eta$ and $\zeta$.
Since we concentrate here on slow fracture with velocities far below the Rayleigh speed, the assumption of static viscoelasticity is legitimiate, thus
%\begin{equation} \label{model::eq3}
%\frac{\partial \sigma_{ik}}{\partial x_k} = 0.
$\partial \sigma_{ik}/\partial x_k=0$.
%\end{equation}
On the crack contour, the total normal and shear stresses have to vanish, $\sigma_{nn}=\sigma_{n\tangential}=0$, with the interface normal and tangential directions $n$ and $\tangential$.
The driving force for crack propagation is given by the chemical potential \cite{Nozieres}
\begin{equation} \label{steady:eq1}
\mu_s=\Omega(\sigma_{ij}^{(el)}\epsilon_{ij}/2-\gamma\kappa),
\end{equation}
with $\gamma$ being the interfacial energy per unit area and $\Omega$ the atomic volume;
the interface curvature $\kappa$ is positive for a convex crack shape.
%the atomic volume $\Omega$ appears since the chemical potential is defined as free energy per particle.
Surface diffusion leads to the following expression for the normal velocity at each interface point
\begin{equation} \label{steady:eq2}
v_n = - \frac{D}{\gamma \Omega} \frac{\partial^2\mu_s}{\partial \tangential^2},
\end{equation}
with the surface diffusion constant $D$ (dimension $\mathrm{m}^4/\mathrm{s}$).
Notice that $\timescale := 2\eta(1+\nu)/E$ defines a timescale, thus $(D\timescale)^{1/4}$ defines a lengthscale parameter which ultimately leads to selection of the tip scale.
In the general case, another timescale (which does not differ significantly from $\timescale$) is set similarly by the viscous coefficient $\zeta$, and we discuss the specific case that these timescales are equal, i.e. $\zeta = 2\eta [v/(1-2\nu) + 1/3]$.
Of course, this simplification is only relevant for mode I fracture, as the second scale does not appear in mode III.
Altogether, the above set of equations fully defines the problem.

We note that for steady state growth with velocity $v$, the last equation can be integrated once, and we obtain
\begin{equation} \label{steady:eq3}
vy =  \frac{D}{\gamma \Omega} \frac{\partial\mu_s}{\partial \tangential}.
\end{equation}

%\section{Mode III fracture}
%{\em Sharp interface modeling.}
We illustrate the procedure to solve the moving-boundary viscoelastic problem for mode III fracture in the steady state regime;
for mode I loadings a similar approach can be used, which will be explained in detail elsewhere.
The crack is located in the $xy$ plane and propagates in positive $x$ direction with velocity $v$.
Then Newton's equation reads
\begin{equation} \label{mode3::eq5}
\nabla^2\left( u_z + \timescale\dot{u}_z \right) = 0,
\end{equation}
from which we obtain by differentiation
%\begin{equation}
$\nabla^2\sigma_{xz} = \nabla^2\sigma_{yz} = 0$.
%\end{equation}
We represent the total stress through an analytical complex potential $\Sigma$ with $\sigma_{xz}=\Im (\Sigma)$ and $\sigma_{yz}=\Re (\Sigma)$.
For steady state growth, $\sigma_{iz}^{(vis)}=-v\timescale\sigma_{iz, x}^{(el)}$, and we therefore make a similar ansatz for the representation of the elastic fields through an analytical function $\Sigma^{(el)}$, $\sigma_{xz}^{(el)}=\Im (\Sigma^{(el)})$ and $\sigma_{yz}^{(el)}=\Re (\Sigma^{(el)})$.
This also guarantees the integrability of the strain field.
The force balance Eq.~(\ref{mode3::eq5}) is then satisfied for solutions of the complex differential equation
\begin{equation} \label{mode3::eq6}
\Sigma^{(el)} - v\timescale \frac{d}{dz} \Sigma^{(el)} = \Sigma,
\end{equation}
with $z=x+iy$.
%We introduce complex stress functions $\Sigma^{(tot)}$ and $\Sigma^{(el)}$, which are related to stresses according to $\sigma_{xz}=\Im \Sigma$ and $\sigma_{yz}=\Re \Sigma$ (all other stress components vanish in mode III).
%We use the ansatz of functions $\Sigma^{(tot)}$ and $\Sigma^{(el)}$, which are analytic in the entire complex plane apart from the negative real axis, which guarantees force equilibrium $\sigma_{ij, j}=0$ and existence of a defect-free elastic displacement field.
For the total stress we use a multipole expansion with a branch cut along the negative real axis,
\begin{equation} \label{mode3::eq1}
\Sigma = \mu \sum_{m=1}^{M=\infty} A_m z^{1/2-m},
\end{equation}
with real coefficients $A_i$.
The main mode $m=1$ is related to the stress intensity factor, $A_1=K_{III}/\mu(2\pi)^{1/2}$ ($\mu=E/2(1+\nu)$ is the shear modulus).
The other coefficients are adjusted such that the boundary condition on the (extended) crack shape, $\sigma_{nz}=0$, is satisfied.
To this end we minimize the residual stress functional $\int \sigma_{nz}^2 ds$ (integrated along the crack contour) with respect to the expansion coefficients and solve the arising linear problem numerically for a known crack shape.
%this linear problem of a finite crack in an infinite medium can easily be solved numerically for a known crack shape;
We restrict the calculation to a finite number of modes $M$ in such a way that the final result does not change noticeably if the accuracy is increased.
%For steady state crack growth, the total stress is related to the elastic stress by $\Sigma^{(tot)} = \Sigma^{(el)} - v\timescale \Sigma^{(el)}_{,x}$.
Eq. (\ref{mode3::eq6}) can now be solved for each mode, and we obtain
\begin{equation} \label{mode3::eq2}
\Sigma^{(el)} = \mu \sum_{m=1}^{M=\infty} A_m \Sigma_m^{(el)},
\end{equation}
with the recursion relation
\begin{eqnarray}
\Sigma_1^{(el)} &=& \pi^{1/2} (v \timescale)^{-1/2} \exp\left( \frac{z}{v \timescale} \right) \erfc\sqrt{\frac{z}{v \timescale}} \label{mode3::eq3}, \\
\Sigma^{(el)}_{m+1} &=& \frac{1}{(m-1/2)v \timescale} \left[ z^{1/2-m} - \Sigma^{(el)}_{m} \right]. \label{mode3::eq4}
\end{eqnarray}
The integration constant is chosen such that far away from the crack tip the purely elastic behavior is retained.
These expressions can then be used to obtain the strain from Hooke's law, and -- after integration -- the displacement field $u_z$.

The strategy of solution is therefore as follows:
First, for known crack shape, the total stress problem is solved in the spirit of Eq.~(\ref{mode3::eq1}), delivering the coefficients of expansion $A_i$.
They, together with a given crack speed are used to determine the elastic stress field according to Eqs.~(\ref{mode3::eq2})-(\ref{mode3::eq4}).
Then, in the next step, the chemical potential (\ref{steady:eq1}) can be computed using Hooke's law.
Finally, the steady state equation (\ref{steady:eq3}) is a nonlocal and nonlinear relation which is used to determine a new guess for the crack shape and velocity.
With them, the whole procedure is iterated until a self-consistent solution is found.

%At the Griffith point, the propagation velocity must vanish, and therefore all elastic energy is converted into surface energy (no bulk dissipation).
We define a dimensionless driving force
\begin{equation}
\Delta = \Delta_I + \Delta_{III} = \frac{1-\nu^2}{2E\gamma} K_I^2 + \frac{1}{4\mu\gamma} K_{III}^2,
\end{equation}
where we already included the possibility of mixed-mode loading, and $\Delta=1$ is the Griffith point.
From now on, we set $\nu=1/3$.
\begin{figure}
\begin{center}
\epsfig{file=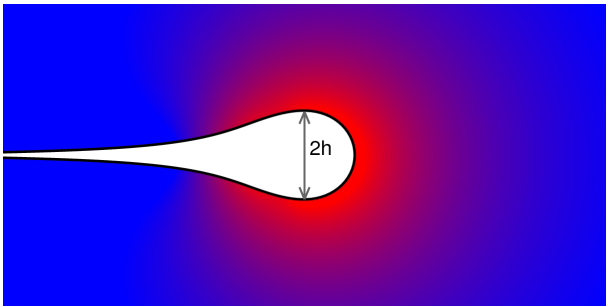, width=8cm} % formerly dissipation.eps
\caption{Shape of a mode III crack for $\Delta=2.5$ in the steady state regime.
The total incoming elastic energy flux is converted into surface energy, surface dissipation and viscous bulk dissipation, which is localized to the scale $v\timescale$ around the crack tip (visualized by the color coding).}
\label{fig1}
\end{center}
\end{figure}
Fig.~\ref{fig1} shows a typical steady state crack shape for mode III loading in the reference frame (Lagrangian coordinates), i.e. the elastic displacement is not included.
First, we clearly see that the crack tip scale is selected self-consistently, and the finite time cusp singularity of the ATG instability does not occur.
Therefore, the presence of viscous bulk dissipation is a way to cure this well-known problem.
Second, it is important that far behind the crack tip the opening decays to zero, which is a consequence of mass conservation, as expressed by the equation of motion for surface diffusion (\ref{steady:eq2}).
Diffusive transport is therefore restricted to the tip region, and no long-range transport is required.
%, and only therefore steady state growth is possible.
Qualitatively, the crack shapes for mode I look very similar.

\begin{figure}
\begin{center}
\epsfig{file=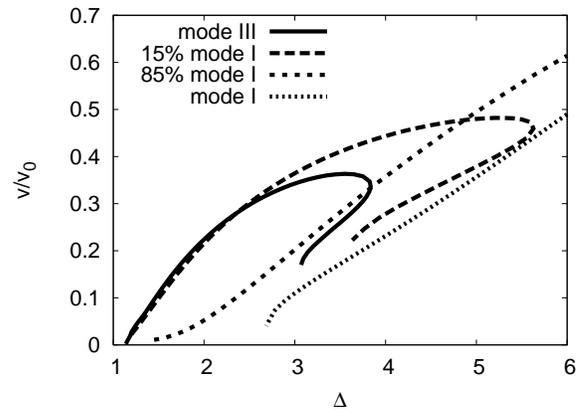, width=8cm} % formerly velocity.eps
\caption{Steady state propagation velocity as function of the driving force for pure mode III and mode I fracture. Additionally, mixed mode situations  with $\Delta_I/\Delta=0.15$ and $\Delta_I/\Delta=0.85$ are displayed. The velocity scale is $v_0=(D\timescale^{-3})^{1/4}$.}
\label{fig2}
\end{center}
\end{figure}
Interestingly, the propagation velocity differs quite significantly for mode I and mode III fracture, as shown in Fig.~\ref{fig2}:
For mode III, the crack speed increases with the driving force, until it reaches a maximum at $\Delta\approx 3.5$, then it decreases, and obviously steady state solutions do not exist beyond the point $\Delta\approx 3.8$, where the stable branch merges with another (unstable) solution.
%We were able to track also this double valued part of the curve by Newton's method to solve the nonlocal steady state equation (\ref{steady:eq2}) with $v_n=v\cos\theta$ ($\theta$ is the angle between the interface normal and the growth direction).
%Careful numerical investigations have shown that the unstable branch finally ends on the horizontal axis, although the accuracy of our multipole expansion method is degraded there.
%%Whereas in the other regimes we see that the inclusion of more modes of expansion does no change the solutions, we cannot predict the behavior quantitatively on the lowest part of the second branch, and therefore it is not shown in the figure.
%However, we point out that this additional steady state branch is unstable anyway, and will therefore not be visible in the fully time-dependent evolution.
Beyond the bifurcation point % $\Delta\approx 3.8$ 
we expect crack branching, in analogy to our findings for fast brittle fracture \cite{Spatschek06}.

\begin{figure}
\begin{center}
\epsfig{file=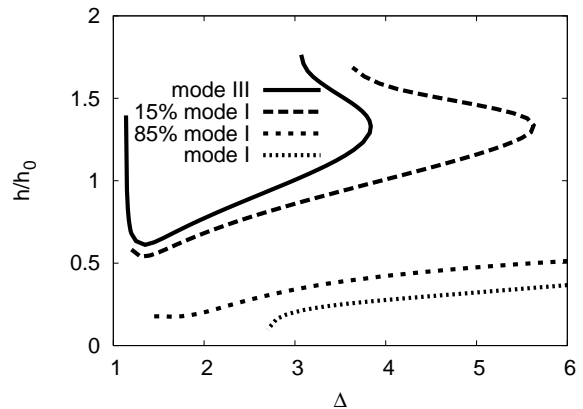, angle=0, width=8cm} % formerly height.eps
\caption{Half crack height as function of the driving force. The lengthscale used here is $h_0=(D\timescale)^{1/4}$.}
\label{fig3}
\end{center}
\end{figure}
Fig.~\ref{fig3} shows the maximum height of the crack as function of the driving force for different loadings.
%Apparently, the lower velocity on the unstable branch of the mode III dominated fracture process is compensated by blunting in order to increase the surface dissipation, and the viscous dissipation becomes smaller.
At $\Delta\approx1.1$ the size of the mode III steady state crack diverges and
$v\to 0$. The viscous dissipation becomes negligible here, but the surface dissipation remains finite.
%and a purely elastic behavior is remained.
%Then, however, we face again the situation of the ATG instability which does not provide a microscopic lengthscale for selection of the tip radius.
This point can be interpreted as  the point of ductile-to-brittle transition:
Below it the size grows indefinitely in time and the crack slows down, while above this point steady state solutions with a finite tip scale exist. 

Starting from a pure mode III crack, we can now include additional mode I loadings.
Fig.~\ref{fig2} shows that this shifts the bifurcation point towards higher values and therefore extends the range of steady state solutions towards higher driving forces.
Again, the crack blunts close to the `nominal' Griffith point $\Delta=1$.
Simultaneously, the propagation velocity is significantly reduced in the regime of small $\Delta$, as can be clearly seen in the comparison between the cases with 15\% and 85\% mode I contribution.
Effectively, this establishes an interval of driving forces, where the crack speed is very low, and only after this plateau it sharply increases;
this effect becomes more pronounced as the crack loading is more mode I dominated. %, and leads to a shift of the `apparent' Griffith point.
The same plateau can also be found in the tip scale, see Fig.~\ref{fig3}.

%Again, the tip scale diverges already slightly above the nominal Griffith point $\Delta=1$, which reflects the fact that the velocity dependent selection mechanism through viscosity cannot predict a tip scale if the crack does not propagate (strictly at the Griffith point) without the presence of an additional microscopic length scale, and which is not contained in the model.

For the case of mode I, finally, steady state solutions do not exist below $\Delta=2.6$;
this result has to be interpreted as a limiting case with very slow creep with velocities and tip radii significantly lower than above the point $\Delta=2.6$.
Literally, of course, growth starts at $\Delta=1$ due to energy conservation.
The presence of this plateau is quite remarkable, as this effectively renormalizes the `apparent' Griffith point -- the driving force where the velocity starts to increase sharply -- to a substantially higher value than $\Delta=1$, although the viscous dissipation remains finite on the `creep branch'.
Again, the crack speed increases monotonically with the driving force, and the bifurcation to unstable growth occurs only at very high driving forces.
%, and at a first glance it seems that here no unstable steady state solutions exist.
%However, the study of mixed mode situations shows that the terminating point of the steady state curve is shifted towards higher values if we add a mode I contribution to a mode III crack.
%Simultaneously, the apparent Griffith point is also shifted.
%This demonstrates that the double-valued part is shifted towards higher driving forces if the fraction of the mode I loading is increased.

\begin{figure}
\begin{center}
\epsfig{file=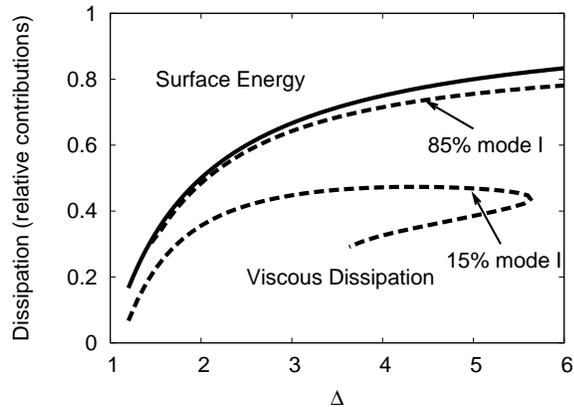, width=8cm} % formerly diss.eps
\caption{Distribution of energy consumption. The fraction above the solid curve is the relative contribution of surface energy generation, the part below the dashed line the viscous dissipation (shown here for two different admixtures of mode I loading; for 15\% mode I also for the unstable branch). The remaining part between the curves is the surface dissipation.}
\label{fig4}
\end{center}
\end{figure}
Finally, Fig.~\ref{fig4} shows how the different mechanisms contribute to the total energy consumption:
From the total dimensionless driving force the amount $\Delta-1$ is dissipated, and the remaining part is used for the creation of the surfaces of the advancing crack.
Obviously, this contribution becomes less important in comparison to the true dissipation for higher driving forces.
The viscous dissipation is
\begin{equation} \label{result::eq1}
\Delta_{v} = \frac{1}{2\gamma v} \int_V R\,dV = \frac{1}{2\gamma} \int_S \sigma_{in}^{(el)} u_{i, x}\, d\tangential
\end{equation}
with the Rayleigh dissipation function $R=\sigma_{ik}^{(vis)}\dot{\epsilon}_{ik}$, and the integration domain $V$ is the solid phase.
The latter equality in Eq.~(\ref{result::eq1}) holds in steady state, where $S$ is the crack contour.
The surface dissipation is
\begin{equation}
\Delta_{s} = \frac{1}{4\gamma} \int_S \sigma_{ik}^{(el)}\epsilon_{ik} n_x d\tangential - 1 = \frac{v}{2D} \int_S y^2 d\tangential,
\end{equation}
with the horizontal component of the interface normal $n_x$;
again, the latter expression, which follows from Eq.~(\ref{steady:eq3}), is valid only in the steady state regime.
Altogether, we have $\Delta=\Delta_s + \Delta_v + 1$.

Starting from the Griffith point the viscous dissipation continuously increases up to the point where the stable steady state solution branch terminates.
%On the unstable branch, the viscous dissipation is always smaller.
%, and the growth of the crack more creep-like.
It is quite remarkable, that for mode I dominated cracks the viscous dissipation $\Delta_v$ is much larger than $\Delta_s$, which shows that bulk dissipation can indeed play a crucial role.
%Since the propagation of the crack requires surface dissipation, its suppression explains the shift of the apparent Griffith point.
Notice that these (dimensionless) predictions do not depend on model parameters.

%The reason why mode I and mode III behave so differently is related to their completely different elastic properties and not a specific property of the viscoelastic crack model discussed here:
%Far behind the tip viscous dissipation is negligible, and the interface motion therefore entirely determined by elastic fields.
%Here it is quite crucial that the elastic energy density enters into the chemical potential, which in particular also contains the tangential stress; the other stress components vanish by boundary conditions.
%Whereas the tangential stress decays fast in mode I (which can be seen from the singular fields), this is not true for mode III, and therefore an equilibrium crack shape does not exist there.
%This can also be understood from the following general argument, that in mode III the two nontrivial stress components can be expressed as real and imaginary part of an analytic function.
%An equilibrium solution would require a vanishing chemical potential along the entire crack shape, which in turn demands a complex potential that is zero there.
%This however, is not compatible with a nontrivial remote stress field.

The obtained results lead to the striking conclusion, that the 
apparent Griffith point may depend quite substantially on the mode of loading.
Although most models in the literature are discussed either in the mode I or mode III case only, we clearly see here that the behavior can be significantly different in these cases, as soon as bulk dissipation is taken into account.
For the specific case of crack propagation in viscoelastic media we obtain that the onset of steady state growth is shifted towards higher values in mode I.
This leads to the interesting consequence that by a rotation of the crack front, which can induce a mode III stress intensity factor, the apparent Griffith threshold can be reduced and the crack speed increased;
%a fully time-dependent three-dimensional simulation is therefore a desirable goal for future research.
a fully time-dependent three-dimensional simulation should shed light on this conjecture.

In summary, we developed a model for crack propagation in viscoelastic media in the spirit of an interfacial pattern formation process.
Motion occurs due to surface diffusion along the extended crack shape, which is -- together with the propagation velocity and the tip scale -- selected self-consistently.
The steady state regime of the model is solved numerically using a series expansion method and a sharp interface description, which efficiently separates the microscopic crack tip scale from the system size.
The results show that the bulk dissipation in the surrounding of the crack tip can play a substantial role especially for higher driving forces, and the crack velocity depends crucially on the mode of loading.

This work was supported by the German Research Foundation under grant SPP 1296 and the German-Israeli Foundation.

\end{document}